%% file: main.tex
\documentclass[conference]{IEEEtran}

\usepackage{hyperref}
\usepackage{authblk}
\usepackage{ulem}
\usepackage{balance}
\usepackage{bm}
\usepackage{listings}
\usepackage{xspace}
\usepackage{xcolor}
\usepackage{colortbl}
\usepackage{multirow}
\usepackage{booktabs}
\usepackage{tcolorbox}
\usepackage{graphicx}
\usepackage{subfigure}
\usepackage{subcaption}
\usepackage{amsmath}
\usepackage{tabularray}
\usepackage{pifont}
\hypersetup{
	colorlinks=true, 
	linkcolor=green,
	filecolor=blue,      
	urlcolor=cyan,
	citecolor=red,
}

\newcommand*{\figu}{{Figure}\xspace}
\newcommand*{\defi}{{Definition}\xspace}

\newcommand*{\sect}{{Section}\xspace}
\newcommand*{\tabl}{{Table}\xspace}

\newcommand{\listitem}[1]{\noindent\textbf{#1}\xspace}
\newcommand*{\tool}{{\textsc{ProtocolGPT}}\xspace}
\newcommand{\header}[1]{{\smallskip \noindent\textbf{#1}}\xspace}

\ifdefined \WithComments
\newcommand{\why}[1]{\textcolor{orange}{why: #1}}
\newcommand{\todo}[1]{\textcolor{red}{todo: #1}}
\newcommand{\clg}[1]{\textcolor{red}{clg: #1}}
\newcommand{\dzj}[1]{\textcolor{purple}{dzj: #1}}
\newcommand{\tai}[1]{\textcolor{green!70!black}{Tai: #1}}
\newcommand\cp[1]{{\color{blue}binpang:~#1}}
\else
\newcommand{\why}[1]{}
\newcommand{\todo}[1]{}
\newcommand{\dzj}[1]{}
\newcommand{\tai}[1]{}
\newcommand\cp[1]{}
\newcommand\clg[1]{}
\fi

\begin{document}


\title{Unleashing the Power of LLM to Infer State Machine from the Protocol Implementation}

\newif\ifanonym
\anonymfalse

\ifanonym
    \author{}
\else
    \author[1]{Haiyang Wei}
    \author[2]{Ligeng Chen}
    \author[3]{Zhengjie Du}
    \author[1]{Yuhan Wu}
    \author[1,4]{Haohui Huang}
    \author[4,5]{Yue Liu}
    \author[5]{\\Guang Cheng}
    \author[1]{Fengyuan Xu}
    \author[1]{Linzhang Wang}
    \author[1]{Bing Mao}
    
    \affil[1]{State Key Laboratory for NovelSoftware Technology, Nanjing University}
    \affil[2]{Honor Device Co., Ltd}
    \affil[3]{Geely Automobile Research Institute}
    \affil[4]{QI-ANXIN Group}
    \affil[5]{School of Cyber Science and Engineering, Southeast University}
\fi

\maketitle

\begin{abstract}
\input{tex/Abstract}
\end{abstract}


\input{tex/Introduction}
\input{tex/Background}

\input{tex/CaseStudy}

\input{tex/ProtocolGPT}
\input{tex/Evaluation}
\input{tex/Conclusion}

\balance
\bibliographystyle{IEEEtran}
\bibliography{ref}

\end{document}

%% file: tex/Abstract.tex
State machines are essential for enhancing protocol analysis to identify vulnerabilities. However, inferring state machines from network protocol implementations is challenging due to complex code syntax and semantics. Traditional dynamic analysis methods often miss critical state transitions due to limited coverage, while static analysis faces path explosion issues. To overcome these challenges, we introduce a novel state machine inference approach utilizing Large Language Models (LLMs), named \tool{}. This method employs retrieval augmented generation technology to enhance a pre-trained model with specific knowledge from protocol implementations. Through effective prompt engineering, we accurately identify and infer state machines. To the best of our knowledge, our approach represents the first state machine inference that leverages the source code of protocol implementations. Our evaluation of six protocol implementations shows that our method achieves a precision of over 90\%, outperforming the baselines by more than 30\%. Furthermore, integrating our approach with protocol fuzzing improves coverage by more than 20\% and uncovers two \textit{0-day} vulnerabilities compared to baseline methods. 


%% file: tex/Introduction.tex
\section{Introduction}
Finite State Machine (FSM) is a core component of protocols, comprising multiple states and transitions based on specific inputs. FSM inference is the fundamental building block of various application domains, including vulnerability detection, software engineering, and network protocol analysis. Improper protocol implements can lead to serious security issues, potentially leading to vulnerabilities~\cite{pham2020aflnet}. So accurately inferring protocol FSMs is crucial for understanding functionality, verification, and finding vulnerabilities\cite{fiterau2020analysis}\cite{bhargavan2017verified}\cite{ba2022stateful}.


Most protocol implementations follow the specifications of network protocol (RFC, i.e., Request for Comments), which define the standard communication functions, input formats, and FSMs~\cite{abdullah2024hermes}. In recent years, researchers have used natural language processing (NLP) techniques to infer FSMs from these specifications~\cite{pacheco2022automated}\cite{abdullah2024hermes}. However, different developers may have discrepant interpretations on the same RFC, resulting in discrepancies between the implemented FSM and the RFC. Even different implementations of the same protocol may have significantly discrepant FSMs, and these discrepancies are often hotspots for vulnerabilities. Therefore, FSMs extracted from specific protocol implementations are more accurate than those in RFCs, more important for protocol security analysis, and can unlock greater potential.

Research on inferring FSMs from protocol implementations can be categorized into two principal types: static analysis~\cite{shi2023extracting} and dynamic analysis~\cite{kleber2018nemesys}~\cite{ye2021netplier}~\cite{kleber2020message}~\cite{gopinath2020mining}. Static analysis approaches, such as control flow analysis, and data flow analysis, infer FSMs by deeply understanding the code structure, eliminating the need for program execution. While static analysis provides deep insights without running the code, it struggles with complicated code structures and uncertain dynamic behavior contexts, often resulting in the path explosion problem. Conversely, dynamic analysis extracts FSMs by tracing program behavior and monitoring runtime data during execution. Although dynamic analysis is beneficial for capturing real-time program behavior, it may overlook critical state transitions due to limited coverage, and its effectiveness significantly depends on the quality of test cases. These limitations underscore the need for a scalable and comprehensive method to infer FSMs from protocol implementations.

Recent studies have shown that LLM has extensive potential for applications in program analysis, which can assist developers with tasks such as programming, code repair, vulnerability detection~\cite{xia2024fuzz4all}~\cite{deng2024large}, even guiding protocol fuzzing~\cite{meng2024large}. However, applying LLM to infer protocol FSM from source code implementations still encounters several challenges: \ding{182} The protocol implementations are highly complex and with great code amount. The core components should be precisely paid attention for analysis, while the irrelavant parts should be excluded preventing from side effects.
\ding{183} The tokens of implementations significantly exceed the input size of LLMs.
Effectively leveraging the knowledge embedded within the repositories to enhance LLM remains a substantial challenge. \ding{184} The procedure for the FSM inference of the protocol is complicated, due to the diverse appearance of static and dynamic status, which cannot be accomplished within a step.
\todo{challenges and tech design should be paired}
To tackle these challenges, we propose a novel perspective on inferring FSMs from protocol implementations by augmenting LLMs with source code. Initially, we isolate FSM-related code from implementations to eliminate irrelevant analysis. The extracted code is then partitioned into appropriately sized segments. Then these segments are converted into a vector store, which augments the pre-trained LLM. Finally, we employ chain-of-thought (CoT)~\cite{cot2022NEURIPS} and background-augmented prompting (BAP)~\cite{luo2023augmentedlargelanguagemodels} techniques to guide the augmented LLM in inferring protocol states, message types, and state transitions.

Our proposed approach aims to overcome several prevailing obstacles in inferring protocol FSMs, offering distinct advantages over traditional techniques such as dynamic program analysis, which typically depends on network traffic samples, and static analysis, which is generally limited by the situation mentioned above. Moreover, our method effectively infers the authentic FSMs from protocol implementations, enabling the identification of discrepancies between the actual protocol implementations and their respective RFC specifications.

To evaluate our approach, we conducted inferences on the FSMs of six protocols (IKEv2, TLS1.3, TLS1.2, BGP, RTSP, and L2TP). The evaluation results demonstrate that \tool{} achieves a over 30\% improvement in precision and a over 55\% enhancement in recall compared to the baselines. In the inferred FSMs, we observed discrepancies between implementations of the same protocol. Additionally, the fuzz tester enhanced by \tool{} achieved over 20\% higher code coverage and discovered two zero-day vulnerabilities. Our contribution can be summarized as follows:


\begin{itemize}
    \item We propose a novel method to enhance LLMs with protocol implementation knowledge, effectively addressing the limitations faced by LLMs in source code analysis.
    \item We firstly introduce an approach for inferring protocol FSMs via implementation with augmented LLMs, enabling the identification of protocol states, message types, and state transitions from protocol implementations.
    \item We developed \tool{}, achieving over 90\% precision in FSM inference. This tool reveals discrepancies in FSMs across discrepant implementations of the same protocol. Additionally, when integrated with AFLNet, \tool{} improves code coverage by over 20\% and discover two 0-day vulnerabilities which are never detected by the others before.
\end{itemize}

%% file: tex/Background.tex
\section{Background and Related Work}\label{sec:background}
In this section, we introduce inference techniques of the protocol's state machine, from rule-based to learning-based.

\subsection{Rule-based Protocol State Machine Inference}
The protocol state machine is a model used to describe the logic and behavior of interaction between peers. Conventional efforts to infer protocol state machine include static analysis~\cite{shi2023extracting} and dynamic analysis \cite{kleber2018nemesys}~\cite{ye2021netplier}~\cite{kleber2020message}~\cite{gopinath2020mining}. 

Static analysis involves directly examining source code using formal methods such as symbolic execution or data flow analysis in protocol reverse engineering. For instance, Shi et al. use abstract interpretation to extract protocol formats from parsers within protocol implementations \cite{shi2023extracting}. However, static analysis methods are limited in terms of generality and scalability, often facing state explosion issues when applied to complex programs. Furthermore, static analysis is not applicable when the source code is unavailable. In contrast, dynamic analysis methods do not require access to the source code. These methods examine data or control flow during message parsing, leveraging previously captured network traffic and applying techniques such as statistical methods and active learning to infer state machines. For instance, Kleber et al. proposed a strategy for classifying message types in binary protocols \cite{kleber2020message} using a similarity metric based on continuous value ranges. Ye et al. introduced a probabilistic approach based on joint distribution for reversing network protocols \cite{ye2021netplier}. Nevertheless, all of these methods share the common limitation of limited coverage.

\subsection{Learning-based Protocol State Machine Inference}
In recent years, machine learning has gained significant attention for its potential to automate and enhance the process of protocol state machine inference~\cite{pacheco2022automated} ~\cite{pandita2013whyper} ~\cite{wong2015dase}. Unlike traditional methods, which often rely on manually crafted rules or heuristics, machine learning approaches leverage data-driven models to learn patterns from observed protocol behaviors. Pacheco et al. proposed a deep learning model based on a natural language data corpus to infer protocol state machines~\cite{pacheco2022automated}. However, these data-driven approaches cannot be applied to protocol implementations in the same way as program analysis methods. 

Large language models are a cutting-edge advancement in artificial intelligence, underpinned by deep learning and natural language processing. Recently, LLMs have excelled in software engineering, including code generating ~\cite{du2024evaluating}, code repairing ~\cite{du2024auto}, and vulnerability detection~\cite{he2023lar}. Furthermore, Meng et al. leverage LLMs to enrich the seed corpus and overcome coverage plateau in protocol fuzzing~\cite{meng2024large}. Therefore, LLMs have the potential to deduce state machines from protocol implementations, showcasing their pivotal role in enhancing protocol security and efficiency.

\todo{Repeat with the challenges in introduction?}\why{challenges: Although LLMs demonstrate excellent potential in program analysis, they face the following challenges. Firstly, the output generated by LLMs may occasionally manifest randomness, leading to the creation of non-existent facts. Secondly, in the face of intricate analysis tasks, LLMs are incapable of rendering efficient completions in a singular effort, necessitating incremental guidance via human-formulated prompts. Moreover, in scenarios involving complicated and advanced programs, the precision of LLMs falls short when compared to some conventional analysis methodologies. This discrepancy can largely be ascribed to the constrained scope of LLMs' context windows, which hampers their ability to thoroughly examine highly complicated and voluminous codebases in a singular analytical session.}

%% file: tex/CaseStudy.tex
\section{Motivation and Case Study}\label{sec:case_study}
Although LLMs are powerful, in this section, we use several cases as motivations to demonstrate the reasons and challenges of inferring FSM from the protocol implementations via LLM. 

\subsection{Discrepancies between Protocol Implements}\label{case_differences}

Discrepancies often arise between different protocol implementations from the same RFC, which creates great resistance for inferring FSM from source code.
Opportunity and challenge coexist, which indicates that inferring from source code implements is more useful for real world analysis.
To illustrate this issue, we conduct a manual analysis of four implementations of the IKEv2 protocol, comparing them with RFC. \tabl{}~\ref{tab:impl_diff} provides a summary of the number of states and transitions. To ensure the reliability of our results, two domain experts thoroughly examined both the RFC and the protocol implementations. 

\input{table/impl_diff}

As illustrated in the table, there are marked discrepancies between the state machines defined in the RFC and those observed in the protocol implementations. Several factors contribute to these discrepancies. Firstly, protocol specifications are typically documented in written form, leading to varying interpretations among different developers. This variability can result in significant discrepancies between the protocol implementations and the RFC. Additionally, the official RFC guidelines for communication mechanisms include several open-ended definitions, allowing developers to design applications tailored to their specific scenarios and functional requirements. Furthermore, developers may introduce custom features to meet specific needs, which may not be supported by other implementations. Consequently, there are inherent discrepancies between the RFCs and implementations. The FSMs inferred from implementations, as opposed to those defined in the RFC, can provide better insights for analyzing software behavior. However, as highlighted in \sect{}~\ref{sec:background}, inferring FSMs from implementations remains a challenge.


\begin{tcolorbox}[size=title]
\textbf{Finding 1:} Due to discrepancies between protocol implementations and RFCs, inferring the FSM through source code implementation is more effective for software analysis than RFC-based derivation.
\end{tcolorbox}

\subsection{Limitation of LLM's Input Length}\label{sec:case_context}

LLM is powerful, but LLM heavily relies on the context information~\cite{li2023long}. Even leveraging the source code implementation to infer FSM is better, but the limted length of context of the current LLMs' remains a great challenge.
For example, GPT-4 has only a context window of 8,192 tokens. This input length suffices for majority of NLP tasks, not enough for program analysis. As illustrated in \tabl{}~\ref{tab:protocol_tokens}, the tokens of protocol implementations far exceeds the context window of the GPT-4.

\input{table/Protocol_tokens}

When inferring FSM from implementation using a pre-trained model, the complexity of the code and the sheer volume of files involved can hinder the model from producing accurate results. Protocol implementations typically comprise components such as state machine, configuration, encryption, etc. However, the code related to FSM is often concentrated in specific files. For instance, the message types are usually defined in a structure. Identifying these specific files can narrow the scope of the code.
\begin{tcolorbox}[size=title]
\textbf{Finding 2:} The limitation of context window renders inferring state machine via LLM a challenging endeavor. 
\end{tcolorbox}

\subsection{LLM should be Fine-grained adjusted for FSM inference}\label{case_infer_fsm}

Intuitively, augmented LLMs can deal with more specific tasks than pre-trained models, e.g., preprocessing the input, guiding the model with eticulously crafted prompts. 

LLMPF~\cite{meng2024large} leverages the syntactic information generated by the LLM to guide protocol fuzzing. Rather than utilizing the LLM augmented with specific knowledge, LLMPF directly employs a pre-trained model. 
We compare the pre-trained model with the augmented model by employing a simple FSM inference case. For the pre-trained LLM, we adopt an approach akin to LLMPF, employing prompt engineering to guide the model in generating FSM. In contrast, we utilize a pre-trained large language model augmented with source code to infer the FSM. Initially, the pre-trained model is enhanced with the protocol implementation, followed by the application of the COT and the BAP to guide the model. The methodology for state machine inference using the augmented LLM is elaborated in \sect{}~\ref{sec:protocolgpt}. For the IKE protocol implementation, we selected strongSwan\cite{strongswan}, a widely used solution in the industry.

\input{table/Extract_FSM_LLM}

\tabl{}~\ref{tab:fsm_llm} presents a comparison of the FSMs inferred by the pre-trained model and the augmented model. The detailed experimental methodology is discussed in \sect{}~\ref{sec:evaluation}. \why{inherent randomness present} The result shows that the augmented model outperforms the pre-trained model by 51.52\% in terms of precision. The enhanced LLM demonstrates the capability to generate reliable state information. While the pre-trained LLM can offer some insights related to FSMs, its precision is limited to 43.48\%. This discrepancy may be attributed to the hallucinations exhibited by LLMs when applied to domain-specific tasks.
\begin{tcolorbox}[size=title]
\textbf{Finding 3:} Compared to the pre-trained model, the LLM augmented with source code exhibits a significantly enhanced capacity to generate a more reliable FSM.
\end{tcolorbox}

%% file: table/impl_diff.tex

\begin{table}[!h]
\centering
\scriptsize
\caption{The state and state transition discrepancies between four IKEv2 implementations and the RFC.}
\label{tab:impl_diff}
\begin{tabular}{@{}ccc@{}}
\toprule
\textbf{Projects} & \textbf{States} & \textbf{Transitions} \\ \midrule
\textit{RFC}~\cite{rfc7296}                       & 8               & 17                   \\ 
\midrule
strongSwan~\cite{strongswan}             & 8               & 23                   \\
libopenikev2~\cite{libopenikev2}         & 22              & 65                   \\
Libreswan~\cite{libreswan}               & 22              & 29                   \\
openswan~\cite{Openswan}                 & 12              & 19                   \\ 
\bottomrule
\end{tabular}
\end{table}

%% file: table/Protocol_tokens.tex
\begin{table}[!h]
\centering
\caption{Detail information of benchmark protocols.}
\vspace{-4pt}
\label{tab:protocol_tokens}
\scalebox{1}{
\begin{tabular}{@{}cccc@{}}
\toprule
\textbf{Protocols} & \textbf{Implementations} & \textbf{Version} & \textbf{Tokens} \\ \midrule
IKEv2              & strongSwan\cite{strongswan}            & f994e0a          & 5,556,945         \\
TLS 1.3            & s2n-tls~\cite{s2ntls}                  & 025f3b2          & 2,685,894         \\
TLS 1.2            & s2n-tls~\cite{s2ntls}                  & 025f3b2          & 2,685,894         \\
BGP                & openbgpd~\cite{openbgpd}               & 08b59c1          & 958,486          \\
RTSP               & feng~\cite{feng}                       & d302a1c          & 84,763           \\
L2TP               & openl2tp~\cite{openl2tp}               & be6c288          & 455,270          \\ \bottomrule
\end{tabular}}
\end{table}

%% file: table/Extract_FSM_LLM.tex

\begin{table}[!h]
\centering
\scriptsize
\caption{The comparison between state machines inferred by pre-trained model and augmented model.}
\vspace{-4pt}
\label{tab:fsm_llm}
\scalebox{1}{
\begin{tabular}{@{}cccc@{}}
\toprule
\textbf{Models} & \textbf{States} & \textbf{Transitions} & \textbf{Precision} \\ \midrule
Pre-trained LLM     & 4               & 12                   & 43.48\%            \\
Augmented LLM       & 8               & 20                   & 95.00\%            \\ \bottomrule
\end{tabular}}
\end{table}

%% file: tex/ProtocolGPT.tex
\section{System Design of ProtocolGPT}\label{sec:protocolgpt}

\begin{figure}[h]
  \centering
  \includegraphics[width=1\linewidth]{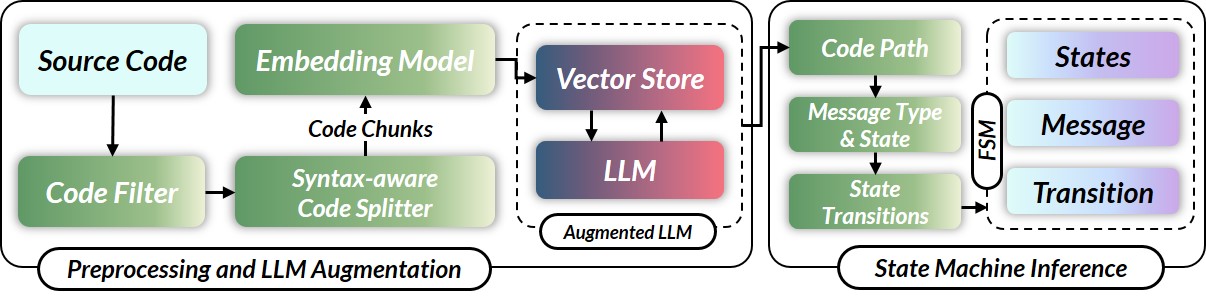}
  \caption{The overview of ProtocolGPT.}
  \label{fig:ProtocolGPT}
\end{figure}

Motivated by the findings in \sect{}~\ref{sec:case_study}, we propose an approach (\tool{}) for inferring state machines based on LLM. The core idea of our approach is to harness the code analysis capabilities of LLM to infer protocol state machines from protocol implementations. \todo{polish}To overcome the challenges aforementioned, we use the augmented LLM that combines the strengths of both retrieval models and generative models. As shown in \figu{}~\ref{fig:ProtocolGPT}, ProtocolGPT consists of two components: LLM augmentation and state machine inference.

\header{LLM Augmentation with Source Code.} 
Initially, we isolate the FSM-related module from the complicated protocol implementations. To guarantee efficient statement inference, the source code is segmented into smaller chunks to bypass limitations of LLM's context window. Moreover, \tool{} maintains the integrity of the original code structure by syntax-aware segmenting for classes, functions, and other structural components. Our method preserves the semantic and contextual information, mitigating potential losses by traditional text segmentation technique. By embedding the LLM with these code chunks, ~\tool{} could more precisely retrieve the most relevant code during the state machine inference.

\header{State Machine Inference.}
To enable the augmented LLM to infer FSM, we integrate the chain-of-thought~\cite{cot2022NEURIPS} and background-augmented prompting techniques~\cite{luo2023augmentedlargelanguagemodels}. \todo{polish}Leveraging domain-specific expertise, we formulate prompts tailored to the task at hand, which guide the augmented LLM in retrieving code segments that define both messages and states. The augmented LLM subsequently generates the message types and states based on the retrieved code. Finally, using the extracted message types and states, we traverse the entire state space to infer the corresponding transitions.

\subsection{LLM Augmentation with Source Code}\label{sec:LLM_augment}

While LLMs exhibit remarkable proficiency in code analysis, their efficacy diminishes significantly when tasked with analyzing large-scale programs. To address this limitation, \tool{} leverages retrieval-augmented generation (RAG) techniques\cite{lewis2021rag}, which empower LLMs by integrating customizable external knowledge sources, such as code repositories. Although RAG models mitigate the limitation of relying on the training data, their performance deteriorates significantly if the external knowledge is not preprocessed. Specifically, protocol implementations often exceed the LLM's context window or include irrelevant information, leading to side effects on retrieval. Consequently, preprocessing the code is essential for LLM augmentation.


\why{delete this paragraph}{}\why{First, the protocol implementation must be filtered to extract sections relevant to the state machine. The source code is then segmented into chunks to comply with the limitation of context window. The code chunks are then embedded and stored in a vector database, contributing to the enhancement of the pre-trained model's performance by enabling efficient retrieval and preserving contextual relevance. During the inference phase, relevant code chunks are retrieved from the vector database and fed into the LLM, supporting the generation of the state machine.}

\subsubsection{Code Filtering}
To ensure scalability and maintainability, protocol implementations are typically structured into distinct modules, including state machines, configuration, testing, etc. Filtering out irrelevant cod not only reduces the cost of LLM prompts but also enhances the accuracy and efficiency of retrieval. Moreover, the tokens in protocol implementations significantly exceeds the context window of LLM. Therefore, extracting state machine module from the codebase forms the first step of the LLM augmentation.

\header{State machine Module Identification.} We observe that variables, constants, structures, and functions related to the state machine are often named according to terms defined in the RFCs. For instance, the CLIENT\_HELLO struct in s2n-tls\cite{s2ntls} corresponds to the \textit{client hello} message in the RFC. Additionally, keywords such as \textit{state} and \textit{message} frequently appear in these context-specific names. Code filter identifies the directory containing the state machine module within the protocol implementation through regular expression matching. Based on these observation, we define a comprehensive set of keywords related to protocol state machines, derived from both RFCs and expert knowledge. 
\why{Add a description of the extensibility of code filtering}The keyword set comprises the message types and states as defined in the RFC, supplemented by custom keywords based on expert knowledge. For instance, in the IKE protocol, Security Association (SA) appears frequently since the state of the SA essentially represents the state of the peer. These specially keywords facilitate the identification of the code related to state machine. Furthermore, our keyword set can be extended to accommodate additional protocols with negligible engineering effort, as these keywords can easily be found in the RFC section headings.

This pre-constructed keyword set is employed to match regular expressions in each document of the protocol implementation, and matching documents are marked as state machine-related code. The subdirectory where the documents have the highest match rate is chosen as the state machine module, providing domain knowledge for LLM augmentation.

\subsubsection{Syntax-aware Code Segmentation}\label{sec:codesplitting}
For the RAG model, the retrieved data is directly provided as context to the prompt, enabling the LLM to generate a response. To retrieve data from the vector database, the documents are transformed into chunks and embeded by the embedding model. However, traditional character-based splitting strategies risk disrupting semantically meaningful elements and may result in the loss of contextual information. Consequently, two critical factors must be taken into account during code segmentation: the chunk size and the syntactic structure of the code.

\header{Appropriate Chunk Size Adjustment.} The input provided to the LLM includes not only the retrieved chunks but also instructions. Appropriately sized chunks contain more precise contextual information, resulting in improved matching accuracy.  However, if the length of the retrieved chunks approaches or exceeds the context window of LLM, the performance of the model is adversely affected. 

\header{Syntactic Program Structure Construction.} Conventional text segmentation overlooks the structural composition of the document. When applied to source code, which is characterized by inherent syntax, these strategies can break apart key components like the classes and functions. Moreover, the rich contextual information embedded within the code may be degraded during the segmentation process.

\begin{figure}[!h]
  \centering
  \includegraphics[width=0.98\linewidth]{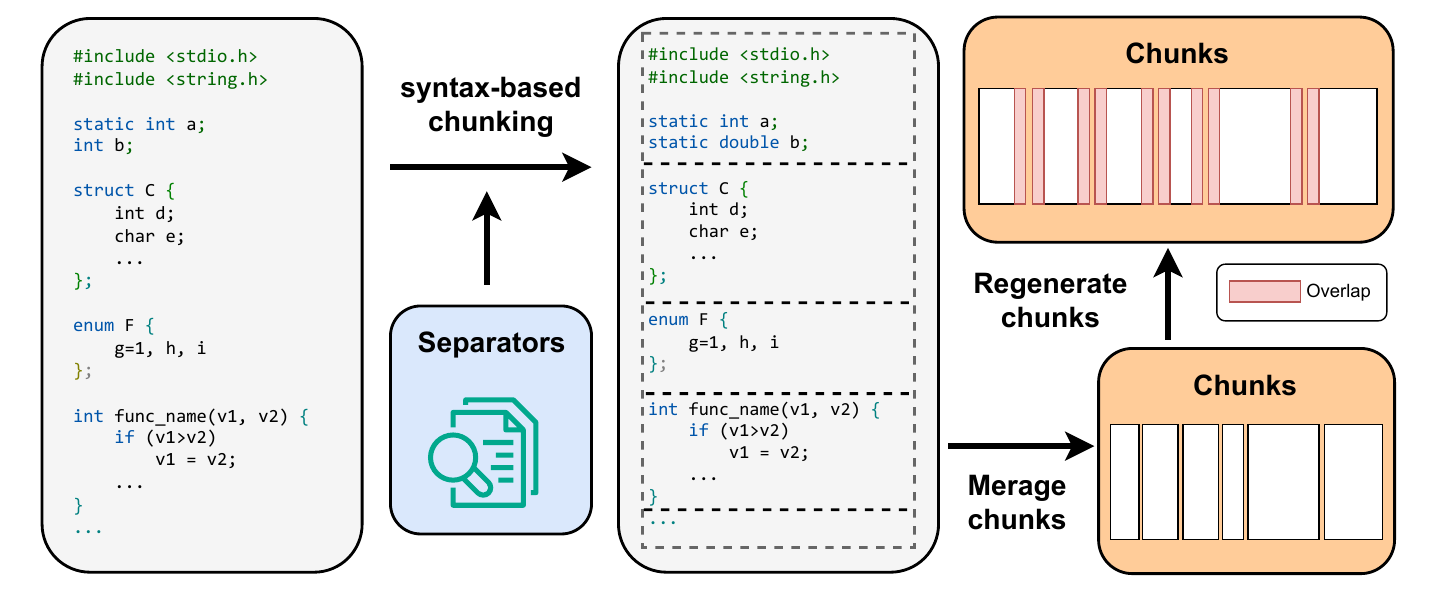}
  \caption{Workflow of Syntax-aware Code Segmentation.}
  \label{fig:syntax_splitting}
\end{figure}

Building upon the preceding observations, we propose a syntax-aware code splitting strategy. This method breaks down code into appropriately sized chunks, ensuring that the syntactic integrity of the code remains intact while also preserving a sufficient level of contextual information. \figu{}~\ref{fig:syntax_splitting} illustrates the workflow of syntax-aware code segmentation. (1) Initially, we define a maximum chunk size, referred to as \textit{MaxChunkSize}. We then recursively decompose the documents using a series of language-specific separators until the size of each chunk is smaller than \textit{MaxChunkSize}. These separators consist of keywords that define critical elements such as structures, classes, functions, and other constructs. For instance, in the C programming language, separators include terms such as \textit{struct}, \textit{class}, \textit{enum}, etc. To facilitate the application of our method across multiple programming languages, we compile a comprehensive list of separators for C, C++, Python, and others. (2) To prevent the creation of overly small chunks that lack sufficient context, we establish a predefined minimum threshold for chunk size, named \textit{MinChunkSize}. If the length of the chunk falls below \textit{MinChunkSize}, it is merged with an adjacent chunk to ensure coherence. (3) Finally, to preserve contextual continuity between chunks, we regenerate new chunks based on previously segmented ones, ensuring a specified degree of overlap is maintained between them.

\subsubsection{Vector Store}
The augmented model employs semantic search to retrieve relevant code from a vector store. These code snippets are then appended to the original prompt, which is subsequently input into the generative model. Retrieval engine is facilitated by converting code into embeddings (high-dimensional vectors). The vector store is specifically optimized for storing embeddings of code, enabling the use of advanced search algorithm to identify matches between similar vectors.

When \tool{} inferring state machine , it initially embeds the prompt to create an embedded version. It then retrieves the vectors from the vector store that most similar to the embedded prompt. Finally, the augmented LLM generate a more contextually relevant response based on the original prompt and the code corresponding to these retrieved vectors. 

For the storage and retrieval of unstructured code, the optimal approach is to embed it and store the resulting vectors. Our approach leverages an embedding model provided by OpenAI\cite{openai} to transform code into vector data. The vector data is subsequently stored and retrieved using a retrieval engine facilitated by FAISS\cite{faiss}. Each vector corresponds to a specific piece of code chunk. The retrieval engine employs the approximate nearest neighbor algorithm to retrieve vectors relevant to the prompt.

\subsection{FSM Inference}\label{sec:4_2_fsm_infer}
Before inferring FSM, it is essential to define the FSM that the model generates. Our primary objective is to examine the transition relationships between different states within the implementation of the protocol. Consequently, we anticipate the FSM generated by the LLM to serve as a high-level model that encompasses both the states and the transitions between these states. Specifically, the FSM details the potential states the protocol server can attain, the pairs of source and destination states, and the conditions necessary for state transitions. To facilitate support for downstream applications, FSM discussed in this paper is assumed to be a non-deterministic FSM that can be transformed into a deterministic FSM. Notably, an non-deterministic FSM may possess multiple initial states, and a single state can transition to several successor states in response to diverse inputs. Inspired by RFCNLP\cite{pacheco2022automated} and StateLifter\cite{shi2023extracting}, we define an non-deterministic FSM in \defi{\textbf{~1}} as a formal model.

\header{Definition 1.} \label{defi:fsm} An FSM is a quintuple $(\Sigma,S,S_{0},E,\delta)$ where
\begin{itemize}
\item \textit{$\Sigma$ is a non-empty set of all message types defined in the protocol implementation.}
\item \textit{$S$ is a non-empty set of protocol states. $S_{0} \subseteq S$ is a non-empty set of all initial states of the protocol application. $E$ is a non-empty set of final states}
\item \textit{$\delta : S \times \Sigma \mapsto 2^{S}$ is a state transition function. A state transition is represented as $\delta(S',m) = S''$, where $\delta$ indicates that when the protocol program receives a message $m$ in state $S'$, it transitions to the next state $S''$.\why{remove the redundant definition of state transiton and add final state}}
\end{itemize}


The FSM inference phase leverages prompt engineering to guide the augmented LLM in deriving FSM from the implementations. However, this process is not instantaneous. As task complexity increases with more steps, the LLM's probability of errors also rises. In contrast, the model demonstrates exceptional accuracy when executing single instructions. To mitigate this limitation, we employ the COT ~\cite{cot2022NEURIPS} to guide the augmented LLM. Specifically, the FSM inference is structured into three stages:
(1) Retrieving the code snippets that define states, messages, and state transitions.
(2) Extracting all the messages and states defined in the protocol implementation.
(3) Identifying all the state transitions and the corresponding message types that trigger these transitions. By following these steps, ~\tool{} gradually extracts the information which forms the FSM of the protocol implementation.

Moreover, the performance of LLMs may be less effective for domain-specific tasks, primarily due to their limited exposure to relevant data. To address this, we employ the BAP~\cite{luo2023augmentedlargelanguagemodels}, which leverages domain knowledge to construct task-specific prompts. \why{Describe details about background-augmented prompt.} Specifically, given the state description $s$ and the downstream task instruction $i$, e.g., "summarize all state transitions", we concatenate $s$ and $i$ to form a prompt that drives the augmented LLM to infer the state transitions.

\begin{figure}[t]
  \centering
  \includegraphics[width=0.95\linewidth]{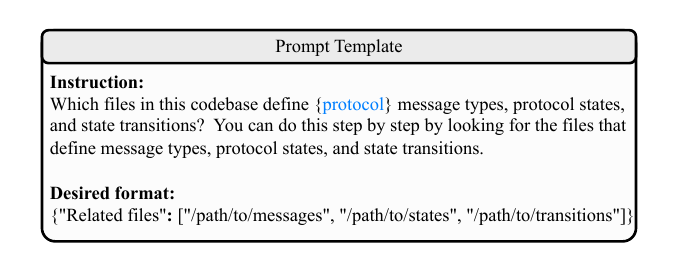}
  \caption{Prompt template for obtaining the code paths related to state machine.}
  \label{fig:prompt_code_path}
\end{figure}

\subsubsection{Code Paths}\label{sec:codepath}
Message types and states are fundamental elements of network protocols. A state transition on the server is triggered when it resides in a particular state and a specific message is transmitted or received. As a result, message types and states are frequently invoked. To improve development efficiency and minimize system complexity, most protocol implementations encapsulate states and message types within defined structures or enumerations. These definitions are typically centralized in dedicated code files. For example, in the TLS protocol implementation s2n-tls, the \textit{message\_type\_t} that defines all supported message types for s2n-tls is located in the file /s2n-tls/blob/main/tls/s2n\_handshake.h. Identifying the relevant documents significantly facilitates the process by which the LLM derives the FSM's states, denoted as $S$ and $\Sigma$.

\figu{}~\ref{fig:prompt_code_path} illustrates the prompt template designed to guide the augmented LLM in identifying the code paths related to state machine. For each protocol implementation, the prompt specifies the protocol type to ensure that the LLM retains all relevant information during generation. 

\begin{figure}[!h]
    \centering
    \subfigure[States] {
            \includegraphics[scale=0.9]{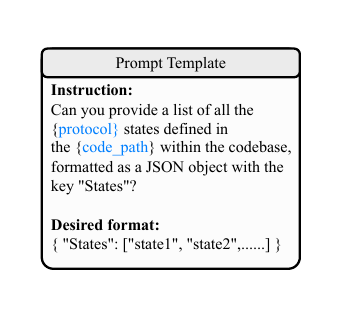}
        }\label{fig:prompt_mes_sta_a}
    \subfigure[Messages] {
            \includegraphics[scale=0.9]{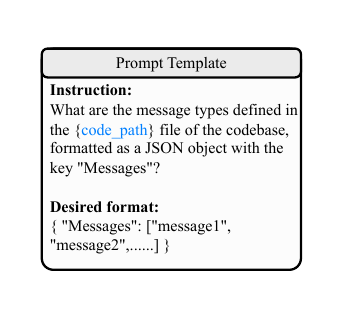}
        }\label{fig:prompt_mes_sta_b}
    \vspace{-8pt}
    \caption{Prompt template used to obtain the states and message types defined in the protocol implementation.}\label{fig:prompt_mes_sta}
\end{figure}



\subsubsection{Message Types \& States}\label{sec:mes_state}
The prompt templates employed to extract the set of message types and states are illustrated in \figu{}~\ref{fig:prompt_mes_sta}. By incorporating previously extracted code paths, these template ensure that the LLM maintains contextual continuity. These prompts guide the augmented LLM in analyzing the previously obtained code files, thereby facilitating the extraction of message types and states defined within the protocol implementation. Fundamentally, our approach leverages the LLM’s code analysis capabilities to extract states and message types from structures or enumerated type variables. It should be noted that our method can infer the FSM from the protocol implementation only when explicit definitions of states and message types are present. Nevertheless, our findings indicate that the Top-50 most widely used open-source protocol implementations provide such explicit definitions, making our approach viable.


\begin{figure}[ht]
  \centering
  \includegraphics[width=0.95\linewidth]{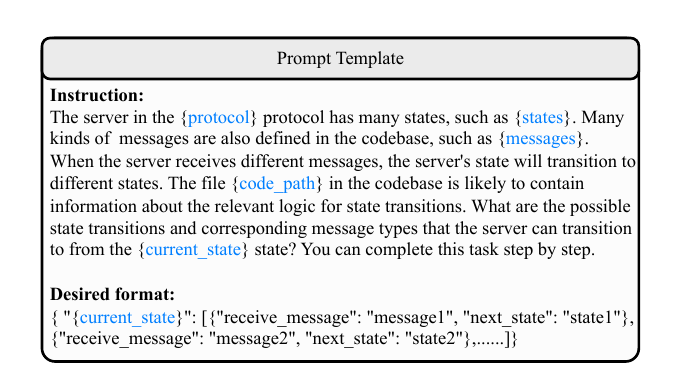}
  \caption{Prompt template for obtaining state transitions contained in protocol implementations.}
  \label{fig:prompt_transitions}
\end{figure}

\subsubsection{State Transitions}
The prompt template employed by ~\tool{} to infer state transitions is depicted in \figu{}~\ref{fig:prompt_transitions}. We guide the augmented LLM to infer state transitions and the corresponding message types by utilizing code paths, predefined states, and message types. Inferring all state transitions within a protocol implementation is inherently a highly complex task. If the LLM were to attempt this task in a single step, it may result in suboptimal performance. Instead, we guide the LLM to systematically infer state transition relationships by iterating through all states. Specifically, \tool{} generates multiple prompts using the extracted states. Then it iteratively replaces the \textit{current\_state} with all previously identified states.

\header{Machine-readable FSM.}
A significant challenge in utilizing LLMs for downstream tasks is ensuring that their outputs are machine-readable. Unlike static and dynamic analysis methods, the outputs of generative model are formulated in natural language. Therefore, without imposing constraints on the output of model, the resulting FSM would be unusable. This challenge can be addressed through fine-tuning prompts, which guide the model to generate outputs according to the given patterns. Specifically, we provide \textit{desired format} in the prompts, which defines the pattern for states, message types, and state transitions. The augmented LLM generates FSM in JSON format, following this specified pattern. For example, each transition includes a \textit{current\_state} key, whose value is an array in ~\figu{}~\ref{fig:prompt_transitions}. Each object in the array contains two properties: \textit{receive\_message} and \textit{next\_state}. The \textit{current\_state} represents the peer's current state, the \textit{next\_state} denotes the destination state, and the \textit{receive\_message} specifies the type of message triggering the transition.

\header{Mitigation of the LLM Hallucination.}
LLMs often generate plausible but non-factual predictions when applied to domain-specific tasks, referred as the hallucination of LLMs\cite{meng2024large}. To mitigate stochastic behavior of the LLM, we conduct 20 iterations of dialogue using the augmented model to extract states, message types, and state transitions. We then select the responses that appear with a probability greater than 80\% across all iterations as the final result.


\subsection{Implementation}
We implement our method with LangChain framework which is a widely adopted open source framework to develop agents based on LLM. The framework provides an extensive suite of interfaces, including Model I/O, Retrieval, and Vector Stores, allowing developers to implement customized functionalities. ~\tool{} implements both LLM augmentation and FSM inference functionalities within the LangChain. During the LLM augmentation, regular expressions are first used to extract state machine module. Documents are then segmented into manageable code chunks by our proposed syntax-aware code segmentation strategy which is implemented based on the text splitter provided by LangChain. The code chunks are then embedded by the model provided by OpenAI and FAISS and stored in a vector store. Ultimately ~\tool{} employs the COT and BAP to guide the augmented LLM to infer FSM. The source code and the experimental data will be released immediately after the work is accepted.

%% file: tex/Evaluation.tex
\section{Experimental Evaluation}\label{sec:evaluation}

To evaluate the effectiveness of ~\tool{}, we try to answer the following research questions:

\listitem{RQ1: }How effective is \tool{} in inferring state machines compared to other methods? (\textit{Section A})

\listitem{RQ2: }How the components effect the performance of \tool{}? (\textit{Section B})


\listitem{RQ3: }What discrepancies can be found between different implementations of the same protocol? (\textit{Section C})

\listitem{RQ4: }How can \tool{} assist find more security bugs? (\textit{Section D})


We select the widely acclaimed GPT-4 model (the latest model at the time of our experiment), renowned for its extensive 1.7 trillion parameters.
Following the recommendation to use a low temperature for generating factual responses, we set it to 0.2. Our experiments were conducted on a machine boasting 32GB of memory and a 12th-Gen Intel(R) Core(TM) i7-12700 CPU, under Ubuntu 20.04.

\header{Benchmark.}
Our benchmark contains 6 network protocol implementations that are widely used in the Internet of Things, spanning a broad range of categories. These protocols vary in format and security levels, covering text-based protocols such as RTSP, binary protocols like IKEv2, encrypted protocols like TLS 1.2, and non-encrypted ones like BGP. For each protocol, we identify the implementation with explicitly defined states on GitHub and select the one with the highest number of stars, as shown in \tabl{}~\ref{tab:protocol_tokens}. 
To set up the ground truth, we recruited three knowledgeable experts with over 72 hours human effort, to independently audit the protocols' repository code and summarize the state machines. Subsequently, the experts collaborated to discuss and refine the proposed state machines, considering a reliable ground truth. 
To minimize bias, the experts we selected for each protocol were highly familiar with the corresponding protocol RFC. Furthermore, we made it clear to each expert that the ground truth should be based solely on objective analysis, excluding any subjective evaluations. Note that manually constructing ground truth from protocol specifications and source code is a common practice in protocol reverse engineering, as demonstrated by prior works such as RFCNLP\cite{pacheco2022automated}, NetPlier\cite{ye2021netplier}, and StateLifter\cite{shi2023extracting}.

\header{Metrics.} 
We focus on the validity of state transitions as it also reflects the validity of the states. State transitions are categorized into four types: $correct$, $partially$ $correct$, $incorrect$, and $not$ $found$. A state transition $T$ is defined as $\{S_i, M, S_t\}$, where $S_i$ is the source state, $M$ is the message triggering the transition and $S_t$ is the destination state. If $S_i$, $M$, and $S_t$ are all correct, then $T$ is classified as $correct$. If one of $S_i$, $M$, or $S_t$ is incorrect, then $T$ is $partially$ $correct$. If more than two elements of $T$ are incorrect, then $T$ is classified as $incorrect$. Undetected state transitions are marked as $not$ $found$. We then respectively compute the precision and recall of the inferred FSM as follows: 
\begin{equation}
\small
Precision = \frac{C}{C + PC + IC}, Recall = \frac{C}{C + PC + NF}
\end{equation}
where $C$, $PC$, $IC$, and $NF$ denote the number of $correct$, $partially$ $correct$, $incorrect$, and $not$ $found$ transitions.

\subsection{Effectiveness}
\input{table/fsm_effectiveness}
To evaluate the effectiveness of \tool{}, we select three popular protocol reverse engineering tools as baselines: RFCNLP\cite{pacheco2022automated}, NetZob\cite{netzob}, NetPlier\cite{ye2021netplier}. RFCNLP, NetZob, and NetPlier represent the state-of-the-art in protocol reverse engineering and have been published in leading academic conferences. \why{GPT-QA}To compare with pre-trained model, we include GPT-QA as an additional baseline. For FSM inference with GPT-QA, no source code hints were provided. \why{polish}Instead, we directly use the prompt in \sect{}~\ref{sec:4_2_fsm_infer} to generate FSMs using GPT-4. Regarding static analysis technique, We fail to find any open-source tools capable of inferring FSMs. Existing static analysis techniques are limited to inferring protocol formats.

\why{First describe the precision and recall, then present case study through state and transitions.} We conduct a \why{how?} analysis of the FSMs inferred by ~\tool{}, GPT-QA, RFCNLP, Netzob, NetPlier \why{and ground truth}. \tabl{}~\ref{tab:fsm_effectiveness} shows the precision and recall of the state transitions. \why{polish}All results are the average of 10 runs and statistically significant with maximum p values less than 0.05. We can observe that the average precision and recall of state transitions inferred by ProtocolGPT achieve 91.42\% and 87.09\%, outperforming baselines by more than 30\% and 55\%. \why{summary:the reason, overcome the limitation of other methods.} Our method significantly outperforms GPT-QA, demonstrating its ability to effectively enhance pre-trained models. Our analysis shows that the learning-based approach, RFCNLP, tends to achieve high precision but low recall, as demonstrated by their more than 10\% higher precision than ours in the case of the RTSP protocol. This can be attributed to the strong reliance on the quality of the labeled data. Additionally, the dynamic analysis methods Netzob and NetPlier exhibit a recall of merely 28\%, which is 60\% lower than that of \tool{}. The limited recall is primarily due to their heavy dependence on the diversity of inputs. \why{too redundant}These results highlight that our method not only enhances pre-trained models but also overcomes the limitations of conventional approaches.

\input{table/fsm_state_transitions}
\header{Analysis of State Transitions.}\todo{polish}To conduct a comprehensive analysis of the states and state transitions inferred by the five tools, we statistical the details of the result, as shown in \tabl{}~\ref{tab:fsm_state_transitions}. GPT-QA generates a substantial incorrect transitions and fails to identify many correct ones, whereas \tool{} demonstrates superior performance in this regard. This contrast highlights the potential of LLM augmentation to effectively mitigate hallucinations produced by pre-trained models. Although our approach has a 10\% lower precision compared to the RFCNLP, it is more robust.Specifically, in the case of RTSP, RFCNLP identify 11 correct transitions, while our method identify 12. The superior performance of our method in other protocols can be attributed to ambiguous definitions in the RFCs. Furthermore, the dynamic analysis methods Netzob and NetPlier often miss substantial correct transitions. For instance, in the case of the BGP protocol, Netzob successfully identifies 5 out of 88 transitions, and NetPlier identifies 6. In summary, compared to the baselines, our method demonstrates the ability to accurately infer reliable FSMs from the protocol implementations.


\subsection{Ablation Study}


Our LLM augmentation method incorporates three primary strategies: code filtering, syntax-aware code segmentation, and code embedding. To evaluate the individual contributions of the strategies, we conducted the following ablation study.
Therefore, We develop five variants of \tool{}:
\begin{itemize}
\item GPT-QA: GPT-4 with only the prompt in \sect{}~\ref{sec:4_2_fsm_infer}.
\item V0: Input with original source code via GPT-4.
\item V1: V0 with code embedding.
\item V2: V1 with syntax-aware code segmentation.
\item V3: V2 with code filter.
\end{itemize}

\input{table/ablation}
The results are presented in \tabl{} ~\ref{tab:ablation}. V0 is unable to work on any protocols due to the limitation of model's context window. The performance of V1, V2, and V3 shows a consistent improvement, indicating that all three components make positive contributions. In comparison to GPT-QA, V3 demonstrates a significant improvement of 55.59\% in precision and 62.95\% in recall. These results underscore the critical role that LLM augmentation plays in the FSM inference.

\header{Strategy 1: Vector Store.} Compared to V0, the improvement in V1 is attributed to the ability to retrieve state information from the embedded code. In contrast, V0 is unable to process the entire codebase. However, when comparing V1 with GPT-QA, there was a noticeable decrease in both precision and recall for V1. We attribute this decline to the different dependent knowledge used by V0 and V1. Specifically, the output of V0 was derived from the training data of GPT-4, whereas V1’s augmented LLM relied on specific knowledge drawn from external datasets. 

\header{Strategy 2: Syntax-aware Code Segmentation.} A comparison between V2 and V1 reveals the impact of the syntax-aware code segmentation. Specifically, V2 demonstrated an average increase of 27.77\% in precision and 12.28\% in recall, underscoring the significant benefits of this strategy. 
This improvement is mainly due to the mechanism splitting the code into more manageable units while preserving the integrity of the syntactical structure.
This approach ensures that semantic information is retained, effectively mitigating the risk of meaning loss during the partitioning process. It effectively addresses the limitation of the LLM's context window. 

\header{Strategy 3: Code Filter.} The comparison between V3 and V2 highlights the impact of the code filter. V3 exhibited a substantial increase in precision and recall by 41.02\% and 55.57\%. This improvement is attributable to the code filter's ability to eliminate code that is irrelevant to FSM, thus reducing the search space for the retrieval. By restricting the augmented LLM from retrieving unrelated data, the efficiency of FSM inference is significantly enhanced. However, the effect of the code filter is less pronounced for concise repository. For example, due to RTSP's compact implementation, the minimal discrepancies between the filtered and unfiltered code led to marginal performance gains of V3 over V2.

\subsection{Discrepancies between Protocol Implementations}

\input{table/impl_diff_res}

\begin{figure}[!h]
\centering     
\subfigure[strongSwan]{\label{fig:strongswan}\includegraphics[scale=0.35]{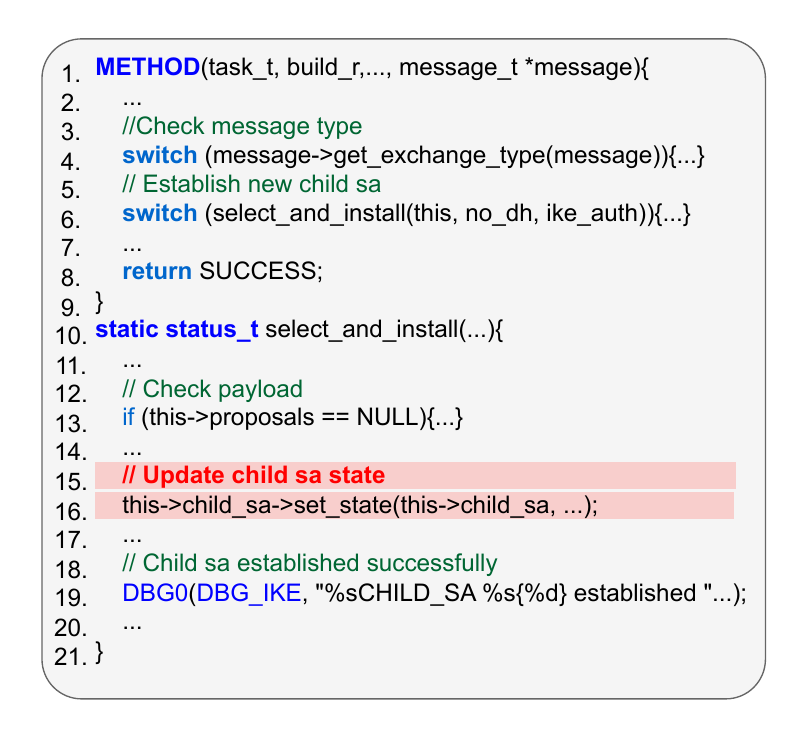}}
\subfigure[libopenikev2]{\label{fig:libopenikev2}\includegraphics[scale=0.35]{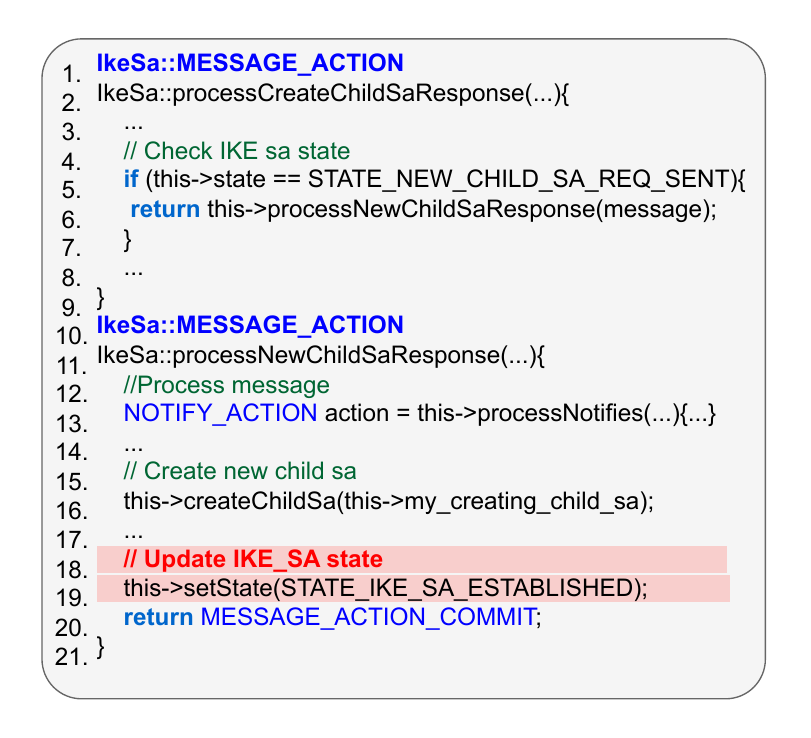}}
\vspace{-8pt}
\caption{State transition differs from the implementations.}
\label{fig:state_differenc}
\end{figure}

To assess \tool{}'s ability to discern variations between different implementations, we applied it to four distinct implementations of the IKEv2 protocol which is a critical component of IPSec suite. A detailed comparative analysis of these implementations is provided in \tabl{}~\ref{tab:impl_diff_res}. Our evaluation revealed discrepancies in the FSMs among the various IKEv2 implementations, highlighting the nuanced discrepancies in their operational behaviors.
For example, in IKEv2, when the communicating parties complete the establishment of an IKE SA, different implementations exhibit varied state transitions upon the creation of a Child SA. \figu{}~\ref{fig:state_differenc} demonstrate the state transition processes for creating a Child SA in strongSwan~\cite{strongswan} and libopenikev2~\cite{libopenikev2}, respectively.

\figu{}~\ref{fig:strongswan} illustrates that in strongSwan, once the peer completes the establishment of the IKE SA and initiates a request to create a Child SA with the other peer, the state of the IKE SA remains unchanged. Instead, this action triggers a transition for the Child SA. In strongSwan, the IKE SA and Child SA manage their state independently. In contrast, as shown in \figu{}~\ref{fig:libopenikev2}, in libopenikev2, after the IKE SA is established, sending a request to create a Child SA results in a transition for the IKE SA to STATE\_NEW\_CHILD\_SA\_REQ\_SENT. Once the Child SA is created, the state reverts to STATE\_IKE\_SA\_ESTABLISHED.

\subsection{Security Applications}
\begin{figure}[!h]
  \centering
  \includegraphics[width=0.8\linewidth]{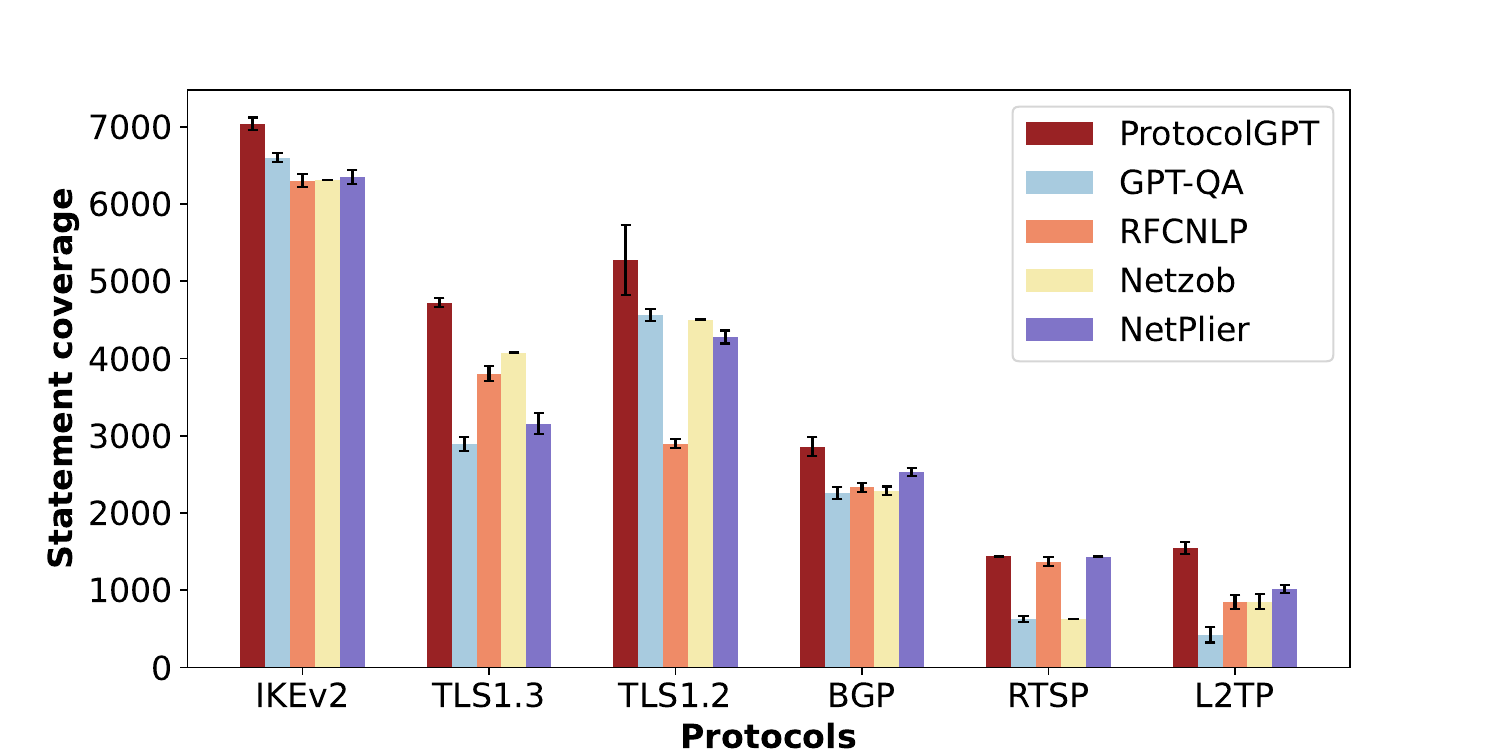}
  \caption{Average line coverage of AFLNet enhanced by different methods in 5 runs of 24 hours, with a 95\% confidence interval.}
  \label{fig:coverage}
\end{figure}

In protocol fuzzing, state machines are instrumental in exploring more code branches and states. AFLNet leverages response codes to construct state machines, enabling targeted testing of the protocol's various states. Utilizing effective message sequences from the corpus as seeds, AFLNet incorporates lightweight mutation algorithms to enhance coverage. By deriving message sequences from the FSMs inferred by ~\tool{}, we generate targeted seeds for AFLNet testing. We adopt the same approach to generate fuzz seeds through the FSMs inferred by GPT-QA\cite{gpt4}, RFCNLP\cite{pacheco2022automated}, Netzob\cite{netzob}, and NetPlier\cite{ye2021netplier}. To reduce randomness in the experiment results, we employed fuzzers across all protocols, running five repetitions over 24 hours.

\header{Code Coverage.} \why{details about improved statement coverage}\figu{}~\ref{fig:coverage} demonstrates that fuzzers enhanced by ~\tool{} outperforms GPT-QA, RFCNLP, Netzob and NetPlier by 31.80\%, 30.39\%, 22.66\%, 21.99\% in terms of statement coverage.
This is attributed to ~\tool{}'s ability to identify a greater number of state transitions, facilitating the discovery of additional code paths. However, in the context of encryption protocols such as IKEv2, TLS1.2, and TLS1.3, AFLNet encounters significant constraints due to its inability to access cryptographic keys during fuzzing, thereby limiting the testing scope to pre-key negotiation phases. The BGP protocol has unique state transitions triggered by events like routing updates. As a result, message-based seeds can not fully cover the code during fuzzing, which highlights the need for event-driven triggers in protocol fuzzing.

\header{Vulnerabilities.}The fuzzer enhanced by \tool{} detects two 0-day vulnerabilities and 18 previously known vulnerabilities. In contrast, other methods detect only five of the known vulnerabilities. Both 0-day vulnerabilities are present in Feng's repository\cite{feng}, requiring the RTSP server to reach a specific state for successful exploitation. The attackers can exploit these vulnerabilities by sending specially crafted requests, causing the server to crash and executing a Denial-of-Service (DoS) attack. Upon submitting the bug report, the developer promptly confirms the two 0-day vulnerabilities. This suggests that the implementation is still active and in use. Consequently, these two 0-day vulnerabilities are likely to present a security threat to industrial software. These detected vulnerabilities can be attributed to our methodology, which facilitates the inference of a reliable protocol FSM, enabling the fuzzer to trigger a wider range of states.

%% file: table/fsm_effectiveness.tex
\begin{table}[!h]
\centering
\scriptsize
\caption{Precision and recall of the state transitions inferred by ProtocolGPT, GPT-QA, RFCNLP, Netzob, and NetPlier.}
\vspace{-4pt}
\label{tab:fsm_effectiveness}
\resizebox{!}{0.15\linewidth}{
\begin{tabular}{@{}c|cc|cc|cc|cc|cc@{}}
\toprule
\multirow{2}{*}{\textbf{Protocols}} &
  \multicolumn{2}{c|}{\textbf{ProtocolGPT}} &
  \multicolumn{2}{c|}{\textbf{GPT-QA}} &
  \multicolumn{2}{c|}{\textbf{RFCNLP}} &
  \multicolumn{2}{c|}{\textbf{Netzob}} &
  \multicolumn{2}{c}{\textbf{NetPlier}} \\ \cmidrule(l){2-11} 
 & \textbf{P} & \textbf{R} & \textbf{P} & \textbf{R} & \textbf{P} & \textbf{R} & \textbf{P} & \textbf{R} & \textbf{P} & \textbf{R} \\ \midrule
\textbf{IKEv2}   & 95.00      & \textbf{82.61}      & 83.33      & 43.48      & \textbf{100.00}        & 17.39      & 16.28      & 30.43      & 82.26      & 47.83  \\
\textbf{TLS 1.3} & \textbf{93.75}      & \textbf{96.77}      & 30.00      & 10.71      & 70.00      & 21.21      & 43.75      & 21.21      & 58.33      & 21.21 \\
\textbf{TLS 1.2} & \textbf{96.67}      & \textbf{93.55}      & 27.27      & 9.68       & 0.00          & 0.00  & 56.25      & 30.00      & 26.31      & 16.67      \\
\textbf{BGP}     & \textbf{94.44}      & \textbf{98.86}      & 51.85      & 60.87      & 44.37      & 76.14      & 50.00      & 21.73      & 54.55      & 26.09      \\
\textbf{RTSP}    & 70.59      & 54.54      & 14.81      & 18.19      & \textbf{84.62}      & 50.00      & 54.17      & \textbf{59.09}      & 63.64      & 31.82      \\
\textbf{L2TP}    & \textbf{98.08}      & \textbf{96.23}      & 7.69       & 1.92       & 31.25      & 9.43       & 33.33      & 9.43       & 81.25      & 24.53      \\ \midrule
\textbf{Avg.} &
  \multicolumn{1}{l}{\textbf{91.42}} &
  \multicolumn{1}{l|}{\textbf{87.09}} &
  \multicolumn{1}{l}{35.83} &
  \multicolumn{1}{l|}{24.14} &
  \multicolumn{1}{l}{55.04} &
  \multicolumn{1}{l|}{29.03} &
  \multicolumn{1}{l}{42.30} &
  \multicolumn{1}{l|}{28.65} &
  \multicolumn{1}{l}{61.06} &
  \multicolumn{1}{l}{28.03} \\ \bottomrule
\end{tabular}}
\end{table}

%% file: table/fsm_state_transitions.tex
\begin{table}[!h]
\centering
\tiny
\caption{State machines inferring results. S and T denote the number of states and transitions, respectively. The C, PC, I, and NF represent the \textit{correct}, \textit{partially correct}, \textit{incorrect}, and \textit{not found} state transitions, respectively.}
\vspace{-4pt}
\label{tab:fsm_state_transitions}
\begin{tabular}{@{}c|c|c|c|c|c|c|c@{}}
\toprule
\textbf{Protocols}                & \textbf{Approaches} & \textbf{S} & \textbf{T} & \textbf{C} & \textbf{PC} & \textbf{I} & \textbf{NF} \\ \midrule
\multirow{6}{*}{\textbf{IKEv2}}         & \textbf{\textit{GroundTruth}}          & 8  & 23  & -  & -  & -  & -  \\ \cmidrule{2-8}
                               & \textbf{ProtocolGPT} & 8  & 20  & 19 & 0  & 1  & 4  \\
                               & \textbf{GPT-QA}       & 4  & 12  & 10 & 0  & 2  & 13 \\
                               & \textbf{RFCNLP}      & 3  & 4   & 4  & 0  & 0  & 19 \\
                               & \textbf{Netzob}      & 39 & 43  & 7  & 0  & 36 & 16 \\
                               & \textbf{NetPlier}    & 60 & 62  & 36 & 0  & 26 & 12 \\ \midrule
\multirow{6}{*}{\textbf{TLS 1.3}}       & \textbf{\textit{GroundTruth}}          & 15 & 31  & -  & -  & -  & -  \\\cmidrule{2-8}
                               & \textbf{ProtocolGPT} & 17 & 32  & 30 & 0  & 2  & 1  \\
                               & \textbf{GPT-QA}       & 6  & 10  & 3  & 1  & 6  & 18 \\
                               & \textbf{RFCNLP}      & 9  & 10  & 7  & 0  & 3  & 26 \\
                               & \textbf{Netzob}      & 12 & 16  & 7  & 5  & 4  & 19 \\
                               & \textbf{NetPlier}    & 8  & 12  & 7  & 4  & 1  & 22 \\ \midrule
\multirow{6}{*}{\textbf{TLS 1.2}} & \textbf{\textit{GroundTruth}}         & 17         & 31         & -          & -           & -          & -           \\\cmidrule{2-8}
                               & \textbf{ProtocolGPT} & 16 & 30  & 29 & 0  & 1  & 2  \\
                               & \textbf{GPT-QA}       & 4  & 11  & 3  & 5  & 3  & 20 \\
                               & \textbf{RFCNLP}      & 3  & 0   & 0  & 0  & 0  & 0  \\
                               & \textbf{Netzob}      & 12 & 16  & 9  & 1  & 6  & 21 \\
                               & \textbf{NetPlier}    & 16 & 19  & 5  & 0  & 14 & 25 \\ \midrule
\multirow{6}{*}{\textbf{BGP}}  & \textbf{\textit{GroundTruth}}          & 7  & 88  & -  & -  & -  & -  \\\cmidrule{2-8}
                               & \textbf{ProtocolGPT} & 7  & 90  & 85 & 2  & 3  & 1  \\
                               & \textbf{GPT-QA}       & 6  & 27  & 14 & 3  & 10 & 9  \\
                               & \textbf{RFCNLP}      & 6  & 151 & 67 & 60 & 24 & 21 \\
                               & \textbf{Netzob}      & 6  & 10  & 5  & 0  & 5  & 83 \\
                               & \textbf{NetPlier}    & 8  & 11  & 6  & 0  & 5  & 82 \\ \midrule
\multirow{6}{*}{\textbf{RTSP}} & \textbf{\textit{GroundTruth}}          & 4  & 22  & -  & -  & -  & -  \\\cmidrule{2-8}
                               & \textbf{ProtocolGPT} & 6  & 17  & 12 & 4  & 1  & 6  \\
                               & \textbf{GPT-QA}       & 6  & 27  & 4  & 0  & 23 & 18 \\
                               & \textbf{RFCNLP}      & 4  & 13  & 11 & 0  & 2  & 11 \\
                               & \textbf{Netzob}      & 17 & 24  & 13 & 3  & 8  & 6  \\
                               & \textbf{NetPlier}    & 8  & 11  & 7  & 0  & 4  & 15 \\ \midrule
\multirow{6}{*}{\textbf{L2TP}} & \textbf{\textit{GroundTruth}}          & 5  & 53  & -  & -  & -  & -  \\\cmidrule{2-8}
                               & \textbf{ProtocolGPT} & 5  & 52  & 51 & 0  & 1  & 2  \\
                               & \textbf{GPT-QA}       & 6  & 13  & 1  & 3  & 9  & 52 \\
                               & \textbf{RFCNLP}      & 4  & 16  & 5  & 0  & 11 & 48 \\
                               & \textbf{Netzob}      & 13 & 15  & 5  & 2  & 8  & 46 \\
                               & \textbf{NetPlier}    & 13 & 16  & 13 & 0  & 3  & 40 \\ \bottomrule
\end{tabular}
\end{table}

%% file: table/ablation.tex
\begin{table*}[!h]
\centering
\scriptsize
\caption{Comparison of the state machines inferred by the four variants of \tool{}.}
\vspace{-4pt}
\label{tab:ablation}
\scalebox{1}{
\begin{tabular}{@{}c|cc|cc|cc|cc|cc|cc|cc@{}}
\toprule
\multirow{2}{*}{\textbf{\begin{tabular}[c]{@{}c@{}}ProtocolGPT\\ variant\end{tabular}}} &
  \multicolumn{2}{c|}{\textbf{IKEv2}} &
  \multicolumn{2}{c|}{\textbf{TLS 1.3}} &
  \multicolumn{2}{c|}{\textbf{TLS 1.2}} &
  \multicolumn{2}{c|}{\textbf{BGP}} &
  \multicolumn{2}{c|}{\textbf{RTSP}} &
  \multicolumn{2}{c|}{\textbf{L2TP}} &
  \multicolumn{2}{c}{\textbf{AVG}} \\ \cmidrule(l){2-15} 
 &
  \textbf{P(\%)} &
  \textbf{R(\%)} &
  \textbf{P(\%)} &
  \textbf{R(\%)} &
  \textbf{P(\%)} &
  \textbf{R(\%)} &
  \textbf{P(\%)} &
  \textbf{R(\%)} &
  \textbf{P(\%)} &
  \textbf{R(\%)} &
  \textbf{P(\%)} &
  \textbf{R(\%)} &
  \textbf{P(\%)} &
  \textbf{R(\%)} \\ \midrule
\textbf{GPT-QA} & 83.33 & 43.48 & 30.00 & 10.71 & 27.27 & 9.68  & 51.85 & 60.87 & 14.81 & 18.19 & 7.69  & 1.92  & 35.83 & 24.14 \\
\textbf{V0}     & N/A   & N/A   & N/A   & N/A   & N/A   & N/A   & N/A   & N/A   & N/A   & N/A   & N/A   & N/A   & N/A   & N/A   \\
\textbf{V1}     & 33.34 & 52.17 & 20.00 & 3.03  & 6.25  & 6.67  & 17.86 & 21.74 & 58.33 & 31.82 & 0.00  & 0.00  & 22.63 & 19.24 \\
\textbf{V2}     & 37.50 & 26.09 & 40.00 & 25.81 & 51.72 & 46.67 & 35.00 & 30.43 & 70.00 & 31.82 & 68.19 & 28.30 & 50.40 & 31.52 \\
\textbf{V3}     & 95.00 & 82.61 & 93.75 & 96.77 & 96.67 & 93.55 & 94.44 & 98.86 & 70.59 & 54.54 & 98.08 & 96.23 & 91.42 & 87.09 \\ \bottomrule
\end{tabular}
}
\end{table*}

%% file: table/impl_diff_res.tex
\begin{table}[!h]
\centering
\scriptsize
\caption{The state machines of four IKEv2 protocol implementations inferred by ProtocolGPT.}
\vspace{-4pt}
\label{tab:impl_diff_res}
\begin{tabular}{@{}c|c|c|c|c|c|c|c|c@{}}
\toprule
\textbf{Implementations} & \textbf{S} & \textbf{T} & \textbf{C} & \textbf{PC} & \textbf{I} & \textbf{NF} & \textbf{P(\%)} & \textbf{R(\%)} \\ \midrule
\textbf{strongSwan}   & 8   & 20 & 19 & 0 & 1  & 4  & 95.00 & 82.61 \\
\textbf{libopenikev2} & 22  & 43 & 32 & 0 & 11 & 22 & 74.42 & 66.67 \\
\textbf{Libreswan}    & 22  & 30 & 29 & 0 & 1  & 0  & 96.67 & 100.00   \\
\textbf{Openswan}     & 12  & 20 & 18 & 0 & 2  & 0  & 90.00 & 94.74 \\ \bottomrule
\end{tabular}
\end{table}

%% file: tex/Conclusion.tex
\section{Conclusion}

In this paper, we introduce a novel state machine inference method utilizing augmented LLMs and perform benchmark experiments. The results demonstrate the potential of LLMs in deriving robust state machines from protocol implementations. Additionally, the use of ~\tool{} allowed for a comparative analysis of state machines across different implementations of the same protocols, revealing discrepancies among them. Ultimately, our findings confirm that the state machines inferred by ~\tool{} can substantially enhance protocol fuzzers by improving code coverage and aiding in the discovery of previously undetected bugs.